\documentclass[11pt]{article}
\usepackage{jhep}
\usepackage[T1]{fontenc}
\usepackage{bm}
\usepackage{graphicx}
\usepackage{float}
\usepackage{caption}
\usepackage{subcaption}
\usepackage[normalem]{ulem}
\usepackage{gensymb}
\usepackage[utf8]{inputenc}
\usepackage{bbold}
\usepackage{soul}
\usepackage{slashed}
 \usepackage{lipsum}
\usepackage[dvipsnames]{xcolor}
\usepackage{hyperref}
\usepackage{cleveref}
\usepackage{bbold}
\usepackage{amsmath}
\hypersetup{colorlinks=false}
\usepackage{fontawesome} 
\usepackage{mathtools}
\usepackage{cancel}

\Crefname{equation}{Eq.}{Eqs.}
\Crefname{figure}{Fig.}{Figs.}

\preprint{UCI-HEP-TR-2021-23, IPPP/21/52}

\newcommand{\SU}[1]{\mathrm{SU}(#1)}

\newcommand{\Sp}[1]{\mathrm{Sp}(#1)}

\newcommand{\com}{\ , \ }

\newcommand{\be}{\begin{equation}}
\newcommand{\ee}{\end{equation}}
\newcommand{\bea}{\begin{eqnarray}}
\newcommand{\eea}{\end{eqnarray}}
\newcommand{\ba}{\begin{aligned}}
\newcommand{\ea}{\end{aligned}}

\newcommand{\SUw}{\mathrm{SU}(2)_{\rm L}}
\newcommand{\U}[1]{\mathrm{U}(1)_{\rm #1}}

\newcommand{\equaref}[1]{Eq.~(\ref{#1})}

\newcommand{\figref}[1]{Fig.~\ref{#1}}

\newcommand{\secref}[1]{Section~\ref{#1}}
\newcommand{\tabref}[1]{Table.~\ref{#1}}

\title{Dark Matter Freeze-out during $\SUw$ Confinement}
\author[a]{Jessica N. Howard,}
\author[b]{Seyda Ipek,}
\author[a]{Tim M.P. Tait,}
\author[c]{Jessica Turner}

\affiliation[a]{Department of Physics and Astronomy, University of California, Irvine, CA 92697 USA}
\affiliation[b]{Department of Physics, Carleton University, Ottawa, ON, Canada}
\affiliation[c]{Institute for Particle Physics Phenomenology, Durham University, South Road, Durham, U.K.}

\emailAdd{jnhoward@uci.edu}
\emailAdd{sipek@physics.carleton.ca}
\emailAdd{ttait@uci.edu}
\emailAdd{jessica.turner@durham.ac.uk}

\date{\today}
\abstract{We explore the possibility that dark matter is a pair of vector-like fermionic $\SUw$  
doublets and propose a novel mechanism of dark matter production that proceeds through the confinement of the weak sector of the Standard Model. 
This confinement phase causes the Standard Model doublets and dark matter to confine into pions. 
The dark pions freeze-out before the weak sector deconfines and generate a relic abundance of dark matter. We solve the Boltzmann equations for this scenario to determine the scale of confinement and constituent dark matter mass required to produce the observed relic density. We determine
which regions of this parameter space evade direct detection, collider bounds, and 
successfully produce the observed relic density of dark matter. For a TeV scale pair of vector-like fermionic $\SUw$  
doublets, we find the weak confinement scale to be $\sim 700$ TeV.}
\begin{document}
\maketitle

\section{Introduction}\label{sec:intro}
The identity of the dark matter and its role in a theory of fundamental interactions remains one of the most pressing open
questions today, and drives a vibrant program of experimental and theoretical research into Physics beyond the
Standard Model (SM) \cite{Bertone:2018krk}. A key property that distinguishes among different possibilities is the nature of the interactions
between the dark matter and the ingredients of the Standard Model, typically characterized by the masses and
couplings of the mediator particles.

An economical choice is to allow the dark matter to transform under the SM's $\SUw$ weak interaction,
repurposing the electroweak bosons of the Standard Model ($W$, $Z$, and $h$) as the mediators. This results in a prototypical weakly interacting massive particle (WIMP), whose abundance in the Universe can be naturally
understood as a result of it freezing out after an initial period of chemical equilibrium with the SM plasma \cite{Cirelli:2005uq}.
While attractive, an $\SUw$-charged WIMP whose abundance is set by freeze-out is highly constrained. The TeV masses favored
by the dark matter abundance often predict signals which are expected to have been visible at colliders \cite{ATLAS:2021kxv,CMS:2021snz}, 
in searches for ambient dark matter scattering with heavy nuclei \cite{XENON:2018voc}, 
and by searches for high energy annihilation products which make their way to the
Earth \cite{MAGIC:2016xys}.
With dominant couplings typically fixed by $\SUw$ gauge invariance, a specific choice of $\SUw$-charged WIMP freezes out
with the correct abundance for only a very narrow range of masses.
While windows of viable parameter space exist (see e.g. Ref.~\cite{Arakawa:2021vih}), many types of $\SUw$-charged WIMPs naively
appear to be excluded as relics whose abundance is determined by freeze-out.

An $\SUw$-charged WIMP typically freezes out at a temperature $\simeq M / 20$, which for an electroweak-sized
mass corresponds to a period of cosmology that is much earlier than Big Bang Nucleosynthesis, and thus during an epoch that is relatively
unconstrained by observational data. At this time, the Universe may deviate dramatically from our extrapolation based on
the SM, due to unforeseen Physics beyond the Standard Model. Indeed, explorations 
of non-standard cosmological histories, including a period of
early matter domination \cite{Hamdan:2017psw}, 
late entropy injection \cite{Gelmini:2006pw}, and modifications of fundamental parameters such as the 
strength of the $\mathrm{SU}(3)$ coupling \cite{Berger:2020maa,Heurtier:2021rko} have all been shown to lead to dramatically different
expectations in the mapping of WIMP parameter space onto its predicted abundance in the early Universe.

\begin{figure}[t!]
\centering

\includegraphics[width=0.9\textwidth]{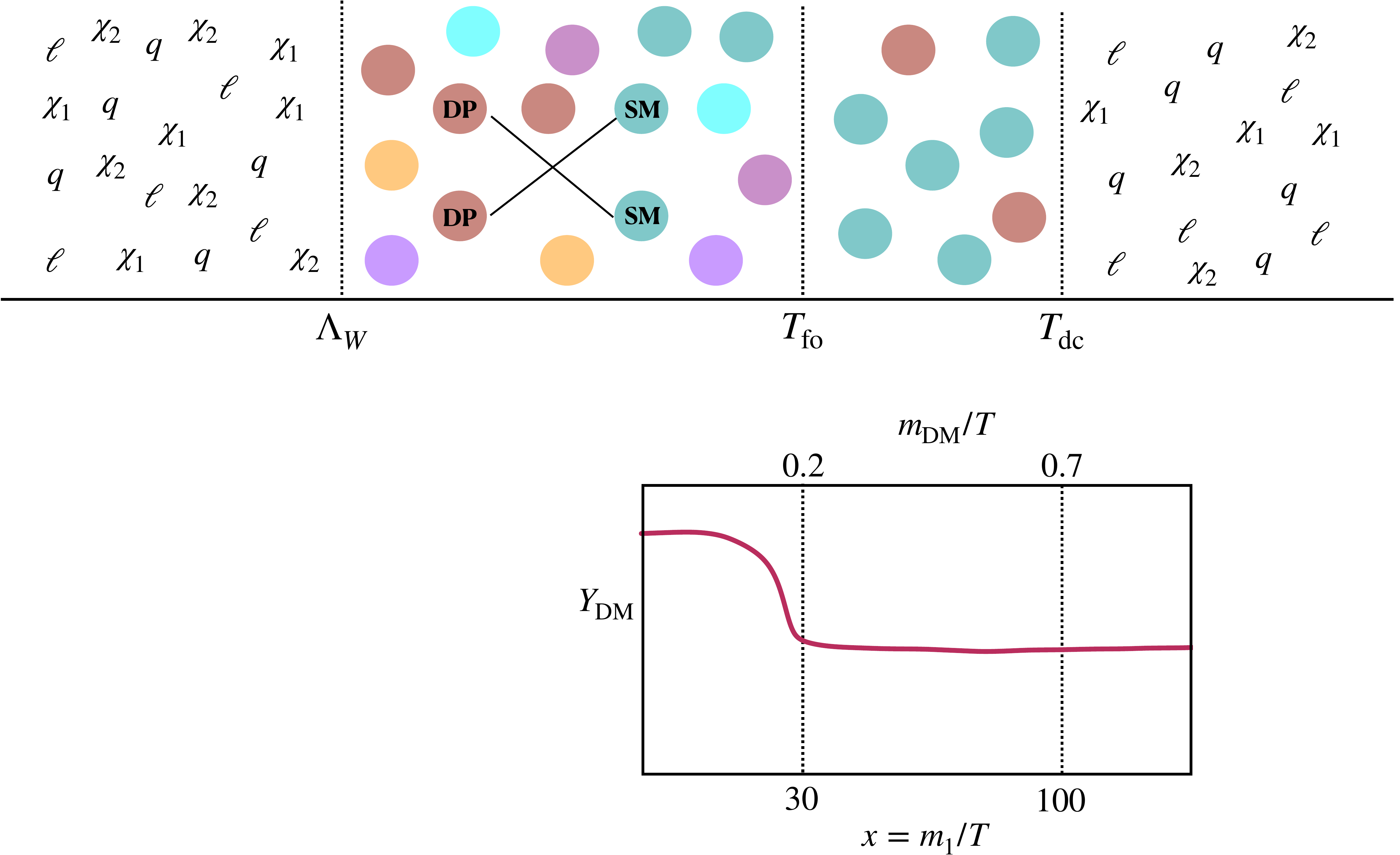}
\caption{Upper panel: A sketch of the cosmological history of the Universe where we assume a period of weak confinement begins at
$\Lambda_W$, at which point the DM ($\chi_1$, $\chi_2$) and SM ($q$, $\ell$) doublets are bound into weak pions. During this epoch,
the freeze-out of dark pions takes place at $T_{\rm fo}$, followed by deconfinement at $T_{\rm dc}$. 
Lower panel: 
The evolution of the dark pion abundance for a representative value of the freeze-out temperature
$x_{\rm fo} = m_1/T_{\rm fo}\simeq 30$, corresponding to a temperature of $0.2m_{\rm DM}$. In our notation, $m_1$ and $m_{\rm DM}$ denote the lightest dark pion and the constituent dark matter masses respectively, see \Cref{sec:dark matterFO} for details.}\label{fig:cosmohistory}
\end{figure}

This article explores a non-standard cosmology that can dramatically change the favored mass range for an $\SUw$-charged WIMP, which makes up the bulk of the dark matter. We introduce dynamics that modify the value of the $\SUw$ interaction strength very early, causing it to confine~\cite{Berger:2019yxb}. This weak confinement causes the left-chiral quarks and leptons of the SM, and a new vector-like  pair of fermionic doublets that plays the role of dark matter, to bind into composite pion-like states that are $\SUw$ neutral. The freeze-out process involves those pions containing
the dark matter annihilating into lighter pions composed entirely of SM fermions. At some time after freeze-out, the $\SUw$
interaction returns to its currently observed value, at which point the pions deconfine, leaving behind the frozen out dark matter.
A sketch of this cosmological history is shown in \Cref{fig:cosmohistory}.

Our work is organized as follows: in \secref{sec:themodel}, we introduce the description of the Universe during an early period of
$\SUw$ confinement, including an additional vector-like pair of fermionic doublets which can play the role of dark matter. In 
\secref{sec:dark matterFO} we discuss the freeze-out process in detail and identify the parameter space leading to the observed
abundance of dark matter today and our results are summarized in \figref{fig:sigmaeff}. The more realistic case including three generations of SM fermions is discussed
in \secref{sec:3gen}.
Finally, we conclude in \secref{sec:outlook} and provide technical details in the appendices.

\section{Weak Confinement and Dark Matter}\label{sec:themodel}

Our dark matter production mechanism involves a temporary cosmological era of $\SUw$ confinement. The possibility that the weak sector was strong in the early universe was initially proposed in \cite{Abbott:1981re, Abbott:1981yg, Claudson:1986ch, tHooft:1998ifg} (see also \cite{Calmet:2000th,Calmet:2001rp,Calmet:2001yd,Lohitsiri:2019wpq}) and the cosmological consequences of such a scenario were studied in Ref.~\cite{Berger:2019yxb}. We refrain from rederiving the complete results of Ref.~\cite{Berger:2019yxb}, which gives a detailed discussion of the gauge and global symmetry breaking patterns as well as the particle content of the confined phase, and instead highlight some key results pertinent for this work:

\begin{itemize}
\item Weak confinement causes the $\SUw$ doublets to condense into bound states analogous to the mesons and baryons of QCD. 
The lowest-lying states are mesons, $\Pi$ and $\eta^\prime$, composed of the SM lepton and quark doublets, $l$ and $q$ respectively. 
These states are contained in the complex antisymmetric scalar field, $\Sigma_{ij}$, 
where $i,j = 1,...., 2N_f$ with $2 N_f$ of left-chiral Weyl fermion fields. For the Standard Model with three generations, $N_f=6$.
\item Following intuition based on chiral symmetry breaking in QCD \cite{Preskill:1980mz,Kosower:1984aw} 
and evidence from lattice simulations, there is a chiral condensate spontaneously breaking the global symmetry:
$\SU{2N_f}\to \Sp{2N_f}$ \cite{Lewis:2011zb,Arthur:2016dir,Karavirta:2011zg,Hayakawa:2013maa,Amato:2015dqp,Leino:2017hgm,Leino:2018qvq}. 
This pattern of symmetry breaking is encoded by the antisymmetric field $\Sigma_{ij}$ acquiring 
a vacuum expectation value $\langle \Sigma_{ij} \rangle = (\Sigma_0)_{ij}$ that satisfies $\Sigma_0^\dagger \Sigma_0 = \Sigma_0 \Sigma_0^\dagger = \mathbb{1}$. 
Neglecting the other SM gauge interactions and Yukawas, this symmetry breaking results in
$2N_f^2 - N_f - 1$ massless Goldstone bosons (GBs) and a single massive pseudo-Goldstone boson (PGB), analogous to the $\eta^\prime$ of QCD. 
\item The dynamics of the confined theory are described by an infrared Lagrangian which is constructed from the scalar field $\Sigma_{ij}$ that contains the massive PGB and massless GBs:
\be\label{eq:sigmaparam}
	\Sigma = 
	\mathrm{exp}\left[i \eta^\prime / \sqrt{N_f} f \right]
	\mathrm{exp}\left[ \sum_{a} 2 i X^a \Pi^{a} / f \right] \ \Sigma_{0}
	\com
\ee 
where $X_{ij}^a$ are the $2N_f^2 - N_f - 1$ broken generators of $\Sp{2N_f}$ and $f$ is the decay constant.
Considering the three SM generations of $\SUw$ doublets, there are 65 massless pions. However, 
loop-induced corrections from the SM gauge and Yukawa interactions 
provide masses to $58$ of the 65 pions.
\item Weak confinement breaks the gauge symmetry of the Standard Model from 
$\SU{3}_{\rm C}\times \U{ \rm Y}$ to $\SU{2}_{\rm C}\times \U{ \rm Q}$, resulting in four massless gauge bosons ($G^{1,2,3}, A^{\prime}$)
and five massive gauge bosons, which can be arranged into a pair of complex gauge bosons (${W^{\prime}}^{\pm}$)
and single real vector boson ($Z^\prime$).
\end{itemize}

We augment the SM particle content by two $\SUw$ doublets, $\chi_1$ and $\chi_2$
(of hypercharges $\pm 1/2$, respectively), which play the role of dark matter. They are assembled into
a pseudo-Dirac state,
\be
{\cal L}_\chi = i \chi^\dagger_1 \bar{\sigma}^\mu D_\mu \chi_1 + i \chi^\dagger_2 \bar{\sigma}^\mu D_\mu \chi_2 + m_{\rm DM} ~\chi_1 \chi_2 + \rm{h.c}\,,
\ee
where $D_\mu$ is a covariant derivative of the unconfined phase and $m_{\rm DM}$ is the mass of the constituent dark matter. This Lagrangian is invariant under a $\U\chi$ symmetry under which $\chi_1$ ($\chi_2$)
are charged $\pm 1$ that ensures their stability.

The infrared Lagrangian, describing the dynamics of the confined theory, has the form
\be\label{eq:potential1}
	 \mathcal{L}_{{\rm IR}}\supset
	 \frac{f^2}{4} \, \mathrm{Tr}\left[ D_\mu \Sigma^\dagger D^\mu \Sigma \right] 
	+ \Lambda_W^3{\rm Tr}[M\Sigma + \Sigma^\dagger M^{T}]
	+ \kappa \Lambda_{\rm W}^2 f^2 {\rm Re}[{\rm det} \, \Sigma] 
	+ \Delta \mathcal{L}\,,
\ee
where $D_\mu $ is a covariant derivative of the confined phase,
$\Lambda_{\rm W} \sim 4 \pi f$ is the weak confinement scale,
$\kappa$ is an $\mathcal{O}(1)$ dimensionless number, and $M$ is the
mass matrix, treated as an $\SU{2N_f}$-breaking spurion in the limit $m_{\rm DM} \ll \Lambda_{\rm W}$. 
In the simplified case where we consider a single generation of $\SUw$ doublets together with the dark matter, $2N_f = 6$ 
and the mass matrix, defined in the basis $\{ \ell, q^{R}, q^{G}, q^{B}, \chi_1, \chi_2 \}$ where $R$, $G$, and $B$ denote the
colors of $\SU{3}_{\rm C}$, is: 
\be
M = \frac{m_{\rm DM}}{2}\begin{pmatrix}
 ~0~ & ~0~ & ~0~ & ~0~ & ~0~ & ~0~ \\
 0 & 0 & 0 & 0 & 0 & 0 \\
 0 & 0 & 0 & 0 & 0 & 0 \\
 0 & 0 & 0 & 0 & 0 & 0 \\
 0 & 0 & 0 & 0 & 0 &1 \\
 0 & 0 & 0 & 0 & -1 & 0 \\
\end{pmatrix}\,.
\ee
 The infrared Lagrangian also contains operators reflecting the explicit breaking of $\SU{2N_f}$ by the gauging of
 $\SU{3}_{\rm C}$ and $\U{Y}$:
\be
\ba
	 \Delta \mathcal{L} & = 
	C_G \Lambda_W^2 f^2 \frac{g_s^2}{16\pi^2} \sum_{a=1,2,3} \mathrm{Tr}[L^a \Sigma^\dagger L^{a T} \Sigma] 
	+ C_A \Lambda_W^2 f^2 \frac{e_Q^2}{16\pi^2} \mathrm{Tr}[Q \Sigma^\dagger Q \Sigma] \\
	&+ C_W \Lambda_W^2 f^2 \frac{g_s^2/2}{16\pi^2} \sum_\pm \sum_{i=1,2} \mathrm{Tr}[L^{i\pm} \Sigma^\dagger L^{i\pm} \Sigma] 
	+ C_Z \Lambda_W^2 f^2 \frac{e_Q^2 / s_Q^2 c_Q^2}{16\pi^2} \ \mathrm{Tr}[J \Sigma^\dagger J \Sigma] \,,
\ea\label{eq:loopcorr}
\ee
where the dimensionless coefficients, $C_G$, $C_A$, $C_W$, and $C_Z$, 
encode the non-perturbative $\SUw$ dynamics, and are expected to be $\mathcal{O}(1)$ ~\cite{Das:1967it,Ayyar:2019exp}. 
The $\SU{2}_{\rm C}$ and hypercharge couplings are denoted as 
$g_s$ and $g^\prime$, respectively, and $\sin\theta_Q = g^\prime/\sqrt{3 g_s^2 + g^{\prime 2}}$ with 
$e_Q\approx g^\prime$ in the limit, $g^\prime \ll g_s$. The generators of the $\SU{2}_{\rm C}$ and $\U{Q}$ are denoted as 
$L^a$ and $Q$, respectively, and $L^{\pm}$ is a combination of $\SU{3}_{\rm C}$ generators which 
couple to the massive vector fields $W^{\prime \pm }$. Finally,
$J$ is a combination of an $\SU{3}_{\rm C}$ and an $\U{Q}$ generator which couple to the massive $Z^\prime$ gauge boson
(see Appendix~\ref{app:A} and Ref.~\cite{Berger:2019yxb} for further details). 

For the remainder of this Section, we consider a simplified toy model consisting of one SM generation of fermionic doublets together with $\chi_1$ and $\chi_2$ (corresponding to $N_f = 3$, for which there are 14 broken generators of the $\SU{6}$ flavor symmetry).
This allows us to extract the most important points in a framework that is simpler to analyze. We return to the more realistic case of three generations plus $\chi_{1,2}$ (corresponding to $N_f = 14$) in \secref{sec:3gen}.

\subsection{Pion Masses and Mass Eigenstates}\label{sec:wpm}

The mass spectrum of the pions during weak confinement is determined from the terms of \Cref{eq:potential1} that are quadratic in the meson fields,
$\mathcal{L}_{{\rm IR}} \rightarrow - (1/2) (M^2_\Pi)_{ab} \Pi^{a} \Pi^b$. Following Ref.~\cite{Berger:2019yxb}, we define $M^2_\Pi$ in the basis $\Pi = \{ \eta', \Pi^a \}$ where $a=1...14$.
In contrast to the case studied in \cite{Berger:2019yxb}, the resulting mass matrix
contains non-diagonal entries mixing the $\eta'$ with the meson dominantly composed of $\chi_1 \chi_2$:
\be
 M_\Pi^2 =  
 \begin{pmatrix}
 M^2_{0,0} & . & . & . & M^2_{0,14}\\
 .& M^2_{1,1} & . & . & .\\
 .& . & . & . & .\\
 .& . & . & M^2_{13,13} & .\\
 M^2_{0,14}& . & . & . & M^2_{14,14}
 \end{pmatrix}\,,
 \ee
and thus the interaction and mass eigenstates are not aligned. We rotate to the mass basis via the unitary transformation $\Pi \rightarrow W \Pi$,
for which
\begin{align}
{M^2_{\rm diag}} = W M_\Pi^2 W^{-1} \,,\label{eq:mmm}
\end{align}
where $W$ is a unitary matrix 
\begin{align}
 W =  
 \begin{pmatrix}
 \cos\theta & . & . & . & \sin\theta\\
 .& 1 & . & . & .\\
 .& . & . & . & .\\
 .& . & . & 1 & .\\
 -\sin\theta& . & . & . & \cos\theta 
 \end{pmatrix}\,,
\end{align}
with
\be \label{eq:mixing}
 \tan 2\theta = 2\frac{M^2_{0,14}}{(M^2_{0,0} - M^2_{14,14})}\,,
\ee
and
 \be
\ba
M^2_{0,0} & =24 \kappa \Lambda_{W}^2+ \frac{2 \Lambda_{W}^3 m_{\rm DM}}{3 f^2}\,,~~~
M^2_{0,14} &= -\frac{2 \sqrt{2} \Lambda_{W}^3 m_{\rm DM}}{3 f^2}\,,~~~
M^2_{14,14} & = \frac{4 \Lambda_{W}^3 m_{\rm DM}}{3 f^2}\,.
\ea\label{eq:msq1}
\ee
Substituting \Cref{eq:msq1} into \Cref{eq:mixing}, we find that
\be
\tan 2\theta = \frac{2 \sqrt{2} \pi m_{\rm DM}}{\pi m_{\rm DM}
-9\kappa f }\approx -\frac{2 \sqrt{2} \pi m_{\rm DM} }{9 \kappa f}+\mathcal{O}\left(\frac{m_{\rm DM}^2}{f^2}\right)\,,
\ee
where we have taken $\Lambda_W = 4\pi f$. Throughout we assume that $m_{\rm DM} \ll f$ 
and this implies that the mixing between $\eta'$ (which we label as $\Pi_0$) and the $\chi_1\chi_2$ (which we label as $\Pi_{14}$) state is small, $\cos \theta \approx 1$ and $\sin \theta \approx \theta$, and this leads to:
\begin{subequations}
\begin{align}
 \Pi_0^{\rm mass} & \approx \Pi_0^{\rm int} + \theta \Pi_{14}^{\rm int}\,,\\
 \Pi_{14}^{\rm mass} & \approx \Pi_{14}^{\rm int} - \theta \Pi_0^{\rm int}\,,
\end{align} 
\end{subequations}
where $\Pi_i^{\rm mass}= \Pi_i^{\rm int}$ for $i = 1,..., 13$. 
The masses of $\Pi_0^{\rm mass}$ and $ \Pi_{14}^{\rm mass}$ are:
\be
\ba
M^2_0 & \approx 384 \pi^2 f^2 \kappa \left( 1+ \frac{\pi m_{\rm DM}}{9\kappa f} + \mathcal{O}\left(\frac{m^2_{\rm DM}}{f^2}\right) \right)\,,\\
M^2_{14} & \approx \frac{256\pi^3}{3} f m_{\rm DM}\left( 1 - \frac{\pi m_{\rm DM} }{9\kappa f} + \mathcal{O}\left(\frac{m_{\rm DM}^2}{f^2}\right)\right)\,.
 \ea
\ee
\Cref{tab:pionMassesInt} shows the approximate masses of the 15 mesons for the one generation SM case,
as well as their representations under the residual $\U{ \rm Q}\times \SU{2}_{\rm C}$ gauge symmetries, in the small mixing limit. 

\begin{figure}[t!]
\centering
\includegraphics[width=0.9\textwidth]{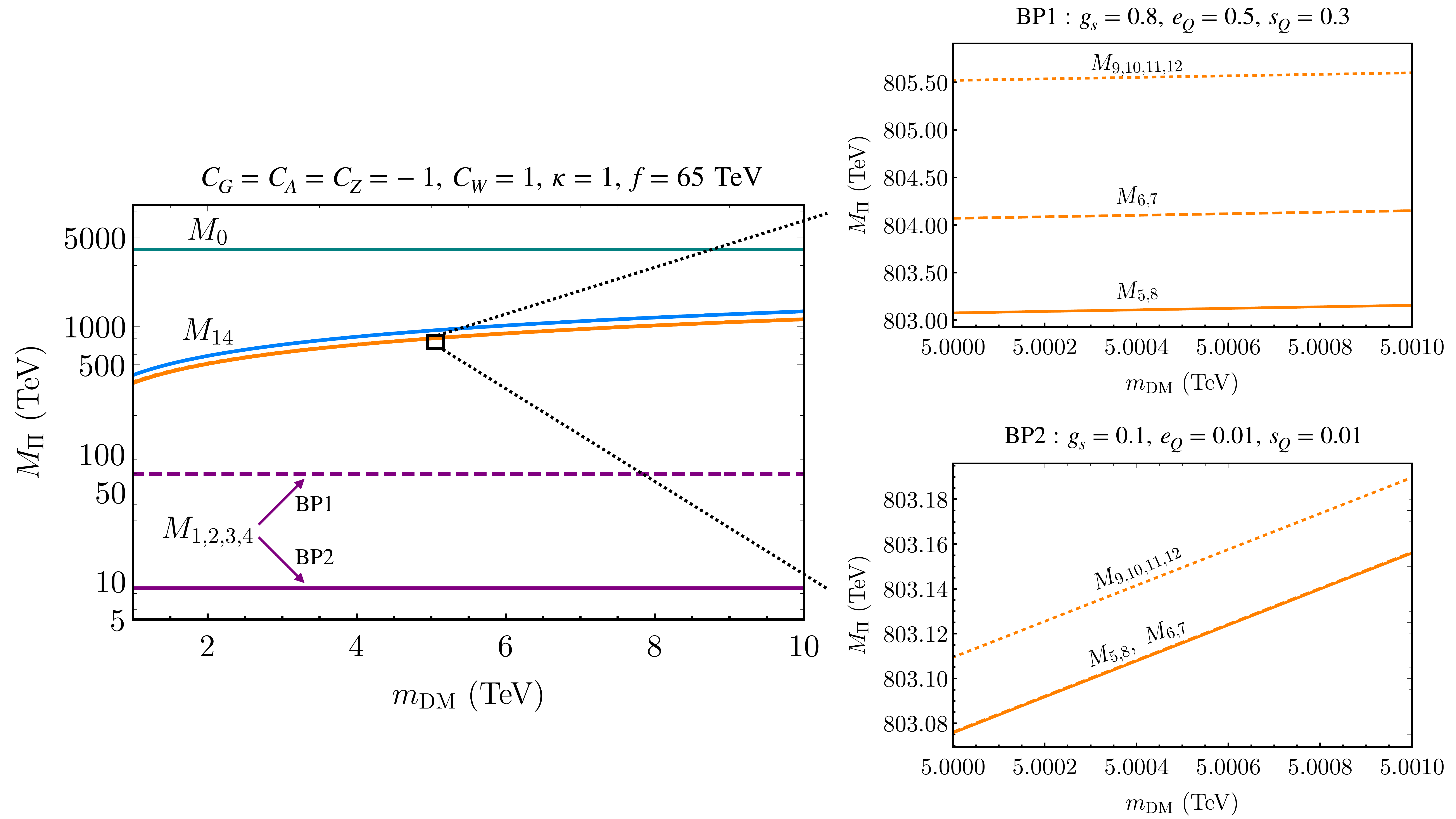}
\caption{Pion masses as a function of $m_{\rm DM}$, assuming $C_G = C_A =C_Z = -1$, $C_W = 1$ and $\kappa=1$, for
two benchmark points: BP1 where $g_s$, $e_Q\simeq g'$ and $s_Q\simeq g'/\sqrt{3g_s^2+g'^2}$ are found by running $g_s$ and $g'$ to $\Lambda_W=4\pi f\simeq800$~TeV;
and BP2 where we take $g_s = 0.1$ and $e_Q=0.01$. $M_{13} = 0$ is not shown.}\label{fig:pionmasses}
\end{figure}

\begin{table}[t!]
\resizebox{\textwidth}{!}{%
\begin{tabular}{ |c|c|c|c|c|} 
\hline
Pion & Mass$^2$ & $\U{Q}$ & SU(2)$_{\rm C}$ & content \\
\hline
\hline
$\Pi^{\rm mass}_{0}$ & $384 \pi^2 f^2 \kappa$ & 0 & 1 & $\chi_{1}, \chi_{2}$\\
\hline
$\Pi^{\rm mass}_{1,2,3,4}$ & $ -\frac{1}{2}C_A e_Q^2 f^2-\frac{3}{2} C_G f^2 g_s^2+C_W
 f^2 g_s^2+\frac{C_Z e_Q^2 f^2}{6 s_Q^2}+\frac{1}{2} C_Z
 e_Q^2 f^2$& $\pm 1$ & 2 & $\ell, q_{D}, q_{S}$\\
\hline
$\Pi^{\rm mass}_{5,8}$ & $64 \pi ^3 f m_{\rm DM}$ & 0 & 1 & $\chi_1, \chi_2, q_{S}$\\
\hline
$\Pi^{\rm mass}_{6,7}$ & $ -2C_A e_Q^2 f^2-2 C_Z e_Q^2 f^2 s_Q^2+\frac{2}{3}
 C_Z e_Q^2 f^2+64 \pi ^3 f m_{\rm DM}$ & $\pm 1$ & 1 & $\ell, \chi_1, \chi_2, q_{S}$\\ 
\hline
$\Pi^{\rm mass}_{9,10,11,12}$ & $ -\frac{1}{2}C_A e_Q^2 f^2-\frac{3}{2} C_G f^2
 g_s^2+\frac{C_Z e_Q^2 f^2}{18 s_Q^2}+64 \pi ^3 f m_{\rm DM}$ & $\pm 1$ & 2 & $\chi_1, \chi_2, q_{D}$\\
\hline
$\Pi^{\rm mass}_{13}$ & $0$ & 0 & 1 & $\ell, q_{S}$\\
 \hline
$\Pi^{\rm mass}_{14}$ & $ \frac{256}{3} \pi ^3 f m_{\rm DM}$ & 0 & 1 & $\chi_{1}, \chi_{2}$\\
 \hline
\end{tabular}
}
\caption{Masses of the pions (for the one SM generation case)
in the small mixing limit, along with their $ \U{Q} \times $SU(2)$_{\rm C}$ charges and constituent $\SUw$ doublet content.}
\label{tab:pionMassesInt}
\end{table}
The specific pion masses depend on the non-perturbative coefficients $C_G$, $C_A$, $C_Z$, $C_W $, and $\kappa$. These could in principle be determined from lattice simulations,
and are expected to be $\mathcal{O}(1)$ based on arguments from naive dimensional analysis \cite{Manohar:1983md}. We proceed under the assumption that
$C_G = C_A =C_Z = -1$ and $C_W = \kappa = 1$.
As is evident from \Cref{tab:pionMassesInt}, the masses of $\Pi_{1,2,3,4}^{\rm mass}$ are independent of $m_{\rm DM}$,
reflecting the fact that they are purely composed of SM quark and lepton doublets, with masses generated via SM gauge interactions, \equaref{eq:loopcorr},
and are typically the lightest of the massive pions. 
The $\Pi_0^{\rm mass}$ is significantly heavier than the other mesons, rendering it unimportant for the freeze-out dynamics due to Boltzmann suppression. We observe that $\Pi_{14}^{\rm mass}$ is $4/3$ times heavier than $\Pi_{5,8}^{\rm mass}$ and hence, we can ignore the effect of $\Pi_{14}^{\rm mass}$ in calculating the dark matter dynamics.
In \Cref{fig:pionmasses}, we show the pion masses as a function of $m_{\rm DM}$ for $f=65$~TeV, corresponding to $\Lambda_W\approx 800$~TeV
(this choice is motivated by discussions of DM abundance in \Cref{sec:dark matterFO}).
We examine two benchmark cases: BP1 where $g_s, g'$ and $s_Q=g'/\sqrt{3g_s^2+g'^2}$ are found by evaluating the running SM coupling constants to approximately $\Lambda_W$ and 
BP2, which is similar to a regime of interest from Ref.~\cite{Berger:2019yxb}. More specifically:
\begin{alignat*}{4}
& \rm{BP1}\quad 
&& g_{s} = 0.8\,,\quad
&& e_{Q} = 0.5\,,\quad 
&& s_Q^2 = 0.12\,, \\
 & \rm{BP2}\, \quad 
&& g_{s} = 0.1\,,\quad
&& e_{Q} = 0.01\,,\quad
&& s_Q^2 = 3.3\times 10^{-3}\,.
\end{alignat*} 
\Cref{fig:pionmasses} indicates that $M_{5,8}, M_{6,7}$ and $M_{9,10,11,12}$ differ slightly due to the loop contributions, and that $M_{5,8}$ are the lightest massive states. 
BP2 has values of $g_s, e_Q$ which are smaller than those in BP1, leading to much smaller
differences between $M_{5,8}, M_{6,7}$ and $M_{9,10,11,12}$, resulting in a more compressed spectrum. 
\subsection{$\U{\chi}$ Eigenstates}
\label{sec:DMnumber}

$\U{\chi}$ remains unbroken during the confined phase, and it is convenient to organize the pions based on their $\U{\chi}$ charges.
This is evident from the fact that the $\U{\chi}$ generator,
\be
Q_\chi=\begin{pmatrix}
 ~0~ & ~0~ & ~0~ & ~0~ & ~0~ & ~0~ \\
 0 & 0 & 0 & 0 & 0 & 0 \\
 0 & 0 & 0 & 0 & 0 & 0 \\
 0 & 0 & 0 & 0 & 0 & 0 \\
 0 & 0 & 0 & 0 & 1 &0 \\
 0 & 0 & 0 & 0 & 0 & -1 \\
\end{pmatrix}\,,
\ee
leaves the vacuum invariant: $Q_{\chi} \Sigma_{0}+\Sigma_{0} Q_{\chi} = 0$.  
To infer the $\U{\chi}$ charges of the pions, we transform $\Sigma$ by an infinitesimal
$\U{\chi}$ rotation:
\be\label{eq:sigma1}
\ba
\Sigma \xrightarrow{\U{\chi}} e^{iQ_\chi \theta_\chi} \Sigma (e^{iQ_\chi \theta_\chi} )^T&\approx 
\Sigma+i \theta_{\chi}\left(Q_{\chi} \Sigma+\Sigma Q_{\chi}\right)+\ldots\,,
\ea
\ee
and expand $\Sigma$ to first order, $\Sigma \simeq \Sigma_{0}+\frac{i}{f} \Pi_{a} X_{a} \Sigma_{0} + ...$, from which
we can extract the transformation of each pion:
\be
\ba
\Pi_{b} \xrightarrow{\U{\chi}}& \Pi_{b} + i\theta_\chi \underbrace{2 \Pi_{a} \text{Tr}[[Q_\chi,X_a],X_b]}_{\delta \Pi_{b}}\,.
\ea
\ee
Using the specific form of the generators $X_a$ and $Q_\chi$ we can explicitly evaluate $\delta \Pi_{a}$ for each $a=0,...14$, and construct complex linear
combinations of pions fields that have definite $\U{\chi}$ charge:
\be \label{eq:mapmtochi}
\ba
 \tilde\Pi^{\pm}_{1} & \equiv & \frac{1}{\sqrt{2}} \left( \Pi^{\rm mass}_{5} \mp i \Pi^{\rm mass}_{8} \right), \\
 \tilde\Pi^{\pm}_{2} & \equiv & \frac{1}{\sqrt{2}} \left( \Pi^{\rm mass}_{6} \mp i \Pi^{\rm mass}_{7} \right), \\
 \tilde\Pi^{\pm}_{3} & \equiv & \frac{1}{\sqrt{2}} \left( \Pi^{\rm mass}_{9} \mp i \Pi^{\rm mass}_{12} \right), \\ 
 \tilde\Pi^{\pm}_{4} & \equiv & \frac{1}{\sqrt{2}} \left( \Pi^{\rm mass}_{10} \mp i \Pi^{\rm mass}_{11} \right),
 \ea
\ee
and $\tilde{\Pi}^0_{i} \equiv \Pi^{\rm mass}_{i}$ for $i \in \{ 0,1,2,3,4,13,14 \} $ 
are left as zero-charge real scalar fields.
Note that these redefinitions commute with the mass basis, as expected.


\subsection{Pion Interactions}
\label{sec:scattering}

The most important interactions of the pions, for our purposes, are four-point vertices arising as residual strong interactions from the confined
$\SUw$ force. These are encoded in the infrared Lagrangian as higher order terms (in powers of $\Pi / f$). Expanding $\Sigma$ to second order:
\be
\ba
 \Sigma(x) &= {\rm exp}\left[ \frac{i \eta'}{\sqrt{N_f} f} \right] {\rm exp}\left[ i \frac{2 \Pi_a (x) X_a}{f} \right] \Sigma_0\\
 &\approx \left[ 1 + i \left(\frac{2 \Pi_a (x) X_a}{f} \right) - \frac{1}{2} \left(\frac{2 \Pi_a (x) X_a}{f} \right)^2 + \mathcal{O}\left(\frac{\Pi^3}{3!f^3}\right)\right] \Sigma_0 \,,\label{eq:sigmaExpansion}
\ea
\ee
where the relevant terms from \Cref{eq:potential1} take the form:
\be
\ba \label{eq:4thOrderLagrangian}
 \mathcal{L}_4 = \frac{4}{f^2} {\rm Tr}_1(a,b,c,d) ~ \Pi_a \Pi_b \partial^\mu \Pi_c \partial_\mu \Pi_d
 + \frac{2 m_{\rm DM}\Lambda_W^3}{3f^4} {\rm Tr}_2(a,b,c,d) ~ \Pi_a \Pi_b \Pi_c \Pi_d \,,
\ea
\ee
with flavor tensors ${\rm Tr}_1$ and $ {\rm Tr}_2$ defined by
\begin{align}\label{eq:Tr1}
 {\rm Tr}_1 (a,b,c,d) &\equiv \frac{1}{4} \Big( {\rm Tr}\left[X_c X_a X_d X_b \right] 
 + {\rm Tr}\left[X_a X_c X_d X_b \right] \Big) \notag\\
 &~~~~~~~ 
- \frac{1}{12} \Big( {\rm Tr}\left[X_c X_a X_b X_d \right] + {\rm Tr}\left[X_a X_c X_b X_d \right] \Big) 
- \frac{1}{3} {\rm Tr}\left[X_a X_b X_c X_d \right]\,,\notag\\
 {\rm Tr}_2 (a,b,c,d) &\equiv - {\rm Tr} \left[ A X_a X_b X_c X_d \right]\,,
\end{align}
where $A \equiv {\rm diag}(\mathbb{0}_{2x2}, ..., \mathbb{0}_{2x2}, \mathbb{1}_{2 \times 2} )$. These expressions are written in the mass
basis, and can be transformed into states of definite $\U{\chi}$ charge via \Cref{eq:mapmtochi}.
 
 Thepions charged under $\SU{2}_{\rm C}$ and $\U{Q}$ will also have gauge interactions with those gauge bosons, contained in the kinetic terms of \Cref{eq:potential1}. However, we have verified that these couplings are small enough at the scales of interest (leading to cross-sections of $\mathcal{O}(10^{-3})$ smaller than those characterizing annihilation into SM pions) that they can be neglected in our freeze-out analysis.

\section{Dark Matter freeze-out}
\label{sec:dark matterFO}

At the time of freeze-out, the dark matter particles are bound into {\em dark pion} (DP) states, and the final abundances of $\chi_{1,2}$ are determined by the frozen-out densities of $\tilde\Pi^{\pm}_{1,2,3,4}$ (each of which contains one $\chi$) and $\tilde{\Pi}^0_0$ and $\tilde{\Pi}^0_{14}$ (each of which contains two $\chi$s). In practice, because of the large mass hierarchy between $\tilde{\Pi}^0_{0,14}$ and $\tilde\Pi^{\pm}_{1,2,3,4}$, it is sufficient to neglect the contributions from the
two neutral states and to focus on the $\U{\chi}$-charged ones.  

The relic abundance of the $\tilde\Pi^{\pm}_{i}$ is controlled by the
temperature, $T_{\rm fo}$, at which their number-changing interactions freeze-out from thermal equilibrium, which in turn depends sensitively on their annihilation cross sections into the lightest neutral pions comprised of SM doublets: $\tilde{\Pi}^+_{i} \tilde{\Pi}^-_{j} \to\tilde{\Pi}^0_{13}\tilde{\Pi}^0_{13}$.
The charged states are typically sufficiently close in mass ($\Delta m/T_{\rm fo}\sim 10^{-2}$) that coannihilation processes
can be important \cite{Griest:1990kh,Servant:2002aq}, and are included in our calculations.
Nonetheless, the relic abundance is dominated by the annihilation of the lightest DP state into the zero-mass SM pion:
$\tilde{\Pi}^+_1 \tilde{\Pi}^-_1 \to \tilde{\Pi}^0_{13} \tilde{\Pi}^0_{13}$.

\subsection{Annihilation Cross-Section}\label{sec:annihilationCX}
\begin{figure}[t!]
\centering
\includegraphics[width=0.9\textwidth]{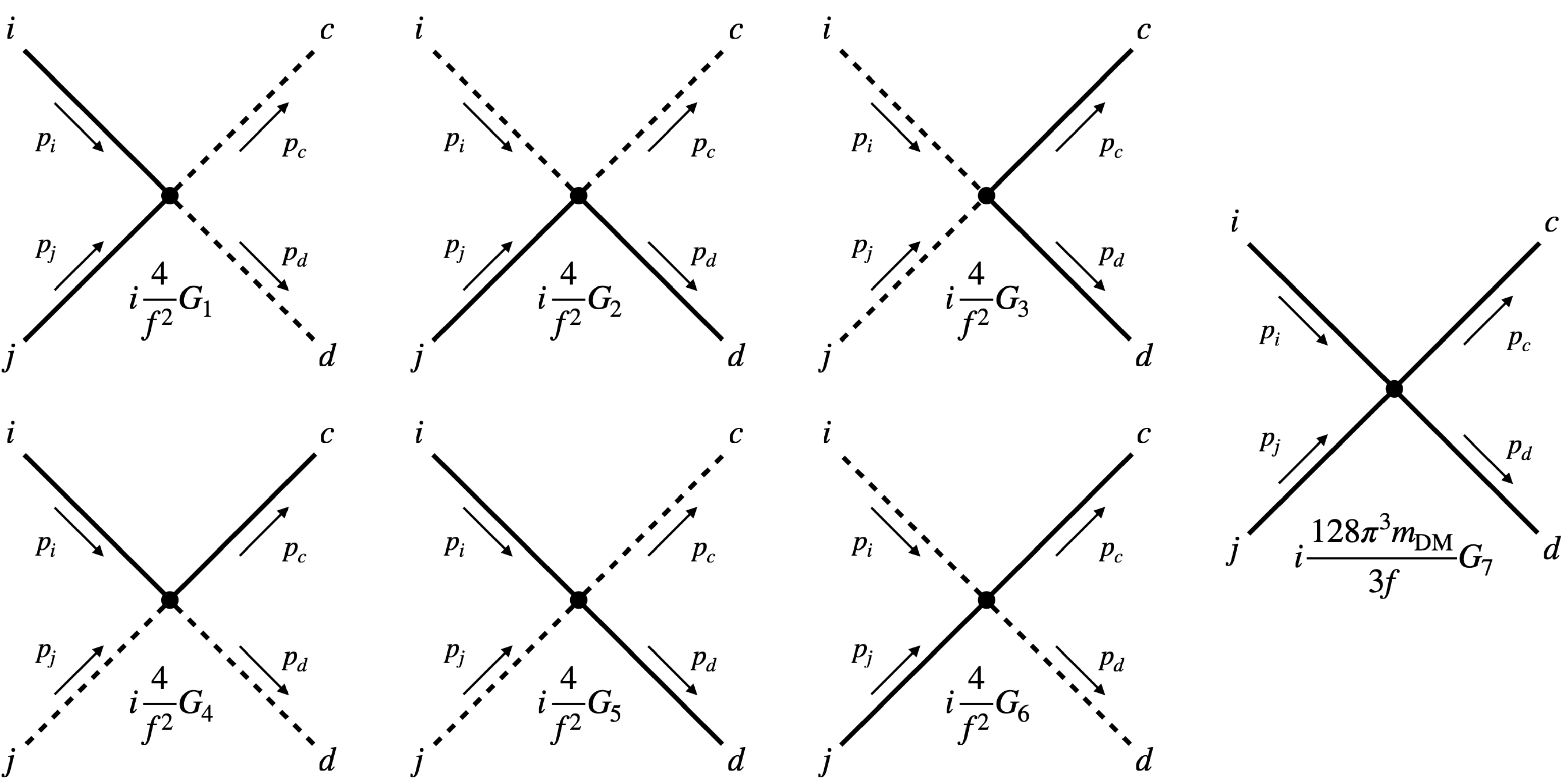} 
\caption{Four-point pion interaction diagrams contributing to the process $\Pi_{a}\Pi_{b}\to \Pi_{c}\Pi_{d}$. The
dashed lines denote fields on which derivatives act, contributing a factor of the corresponding momentum. An incoming (outgoing) field contributes a negative (positive) momentum factor to the matrix element.}\label{fig:IntDiagrams}
\end{figure}
The rate for $\Pi_{i}\Pi_{j}\to \Pi_{c}\Pi_{d}$ is determined by the Feynman diagrams shown in \Cref{fig:IntDiagrams}, where the
dashed (solid) lines indicate legs on which derivatives do (do not) act in the corresponding operator. We define the incoming legs to correspond to the pion flavors
$i, j$, and outgoing to $c, d$. 
The resulting matrix element, $\mathcal{M}$, takes the form
\be
\ba
 i\mathcal{M} = -i \frac{ 4(p_c \cdot p_d)}{f^2}G_1
& + i \frac{4 (p_i \cdot p_c)}{f^2}G_2
 - i \frac{4(p_i \cdot p_j)}{f^2}G_3
 + i \frac{4 (p_j \cdot p_d)}{f^2} G_4\\
 &+ i \frac{4 (p_j \cdot p_c)}{f^2}G_5
 + i \frac{4 (p_i \cdot p_d)}{f^2}G_6
 + i \frac{128\pi^3 m_{\rm DM}}{3f}G_7\,,
\ea\label{eq:ME1}
\ee
where we used $\Lambda_W=4\pi f$ and define:
\begin{alignat*}{4}
& G_1 = {\rm Tr}_1(i,j,c,d)\,, \quad 
&& G_2 = {\rm Tr}_1(d,j,c,i)\,,\quad
&& G_3 = {\rm Tr}_1(c,d,i,j)\,,\quad 
&& G_4 = {\rm Tr}_1(i,c,j,d)\,, \\
 & G_5 = {\rm Tr}_1(i,d,c,j)\,, \quad 
&& G_6 = {\rm Tr}_1(c,j,i,d)\,,\quad
&& G_7 = {\rm Tr}_2(i,j,c,d)\,,
\end{alignat*} 
where ${\rm Tr}_1$ and ${\rm Tr}_2$ are given in \equaref{eq:Tr1}.
Subsequently, the annihilation cross-section can be expressed as,
\be
\ba \label{eq:crossSection}
 \sigma_{ij} (s) = \frac{1}{16 \pi} \frac{1}{\lambda (s, m_i^2, m_j^2)}
 \left[ C_{\rm const} (t_{+}- t_{-}) + \frac{1}{2} C_{\rm lin} (t_{+}^2 - t_{-}^2) + \frac{1}{3} C_{\rm quad} (t_{+}^3 - t_{-}^3) \right]\,,
\ea
\ee
where $\lambda$ is the well-known K\"{a}ll\'{e}n function, $\lambda(x, y, z) \equiv (x - y - z)^2- 4yz$, and 
\be
\ba
 t_{+} &= m_c^2 + m_i^2 - 2 E_c E_i + 2 |\vec{p}_c| |\vec{p}_i|\,,\\
 t_{-} &= m_c^2 + m_i^2 - 2 E_c E_i - 2 |\vec{p}_c| |\vec{p}_i|\,.
\ea
\ee
The above traces define the coefficients inside the square brackets as: 
 \be
 \ba
 C_{\rm const} &\equiv\left(\frac{128\pi^3 m_{\rm DM}}{3f}\right)^2|C|^2 + \frac{4s^2}{f^4} |G_{13,56}|^2 - \frac{2s}{f^2}\left(\frac{128\pi^3 m_{\rm DM}}{3f}\right) \left[C^* G_{13,56} + C G_{13,56}^*\right]\,,\\
 C_{\rm lin} &\equiv \frac{4s}{f^4} \left(G_{13,56} G_{24,56}^* + G_{13,56}^* G_{24,56}\right) - \frac{2}{f^2}\left(\frac{128\pi^3 m_{\rm DM}}{3f}\right)\left[C^* G_{24,56} + C G_{24,56}^*\right]\,,\\
 C_{\rm quad} &\equiv\frac{4}{f^4} |G_{24,56}|^2\,,\\
\ea
\ee
with
\be
\ba
 C &\equiv 
 G_7 + \frac{3m_i^2}{64\pi^3fm_{\rm DM}} (G_2 + G_3 - G_5) + \frac{3m_j^2}{64\pi^3fm_{\rm DM}} (G_3 + G_4 - G_6) \\ 
 & ~~~~~~~~+ \frac{3m_c^2}{64\pi^3fm_{\rm DM}} (G_1 + G_2 - G_6) + \frac{3m_d^2}{64\pi^3fm_{\rm DM}} (G_1 + G_4 - G_5)\,,\\
 G_{13,56} &\equiv G_1 + G_3 - G_5 - G_6\,,\\
 G_{24,56} &\equiv G_2 + G_4 - G_5 - G_6\,.
 \ea
\ee
Note that $64\pi^3 fm_{\rm DM}$ is the mass squared of the lightest pion with DM constituent.
In the non-relativistic limit, the $2\to2$ scattering cross-section can be expanded in terms of the relative velocity, $v = |\vec{v}_i - \vec{v}_j|$, of the incoming particles,
\be
\ba
 \langle \sigma v \rangle = \sigma_0 + \sigma_2 \langle v^2 \rangle + ..
\ea
\ee

At freeze-out, the leading ($s$-wave) term of this expansion dominates over the order higher terms and hence the velocity averaged cross-section is:
\be
\label{eq:thermalavcx}
 \langle\sigma_{ij} v\rangle_{\rm s-wave} = \frac{\lambda^{1/2}(s, m_c^2, m_d^2)}{32 \pi E_a E_b s} 
 \left[ C_{\rm const} + C_{\rm lin} W_1 + C_{\rm quad} W_1^2\right]\,, 
\ee
where
\be
 W_1 = m_i^2 + m_c^2 - \frac{1}{2s}(s + m_i^2 - m_j^2)(s + m_c^2 - m_d^2)\,.
\ee
Further, assuming that the incoming particles are non-relativistic implies that $s = (m_i + m_j)^2$.
For the most significant annihilation processes, $\tilde{\Pi}^+_1 \tilde{\Pi}^-_1 \to \tilde{\Pi}^0_{13} \tilde{\Pi}^0_{13}$, 
$m_c=m_d = 0$ and $m_i^2\simeq m_j^2 \simeq 64\pi^3 fm_{\rm DM} $. In this limit, the parametric dependence of the $s$-wave annihilation cross section is:
\begin{align}
\langle\sigma_{ij} v\rangle_{\rm s-wave} \propto {\rm constant}\times \frac{m_{\rm DM}}{f^3}\,, \label{eq:xsecapprox}
\end{align}
where the overall constant is a combination of various traces and found to be $O(1)$ from numerical analysis. 


\subsection{Freeze-out}
The number density of the dark pions, $n_{\rm DP} = n_{\Pi^\pm_1} + n_{\Pi^\pm_2} + n_{\Pi^\pm_3} + n_{\Pi^\pm_4}$, evolves according the Boltzmann equation \cite{Griest:1990kh}:
\be
\ba
\dot{n}_{{\rm DP}} + 3H {n}_{{\rm DP}} = -\langle\sigma_{\rm eff} v \rangle ( {n}_{{\rm DP}}^2 - {n}_{{\rm DP, eq}}^2)~, \label{eq:DMbolzman}
\ea
\ee
where $H=\sqrt{8\pi^3 g_\ast/90}T^2/M_{\rm Pl}$ is the Hubble rate during radiation domination, 
 ${n}_{\rm DP, eq}= g_* m_1^2 T/(2\pi^2)K_2(m_1/T)$ is the equilibrium number density of the lightest dark pion and $m_{1}$ is the mass of the lightest DP freezing out, more specifically the mass of $\tilde{\Pi}^{\pm}_{1}$. 
The effective co-annihilation cross-section is defined as
\be
\label{eq:sigmaeff}
\ba
\sigma_{\rm eff }&=\sum_{i ,j=1}^{4} \sigma_{i j} \frac{g_{i} g_{j}}{g_{\rm eff }^{2}}\left(1+\Delta_{i}\right)^{3 / 2}\left(1+\Delta_{j}\right)^{3 / 2} e^{-x\left(\Delta_{i}+\Delta_{j}\right)}\,, \\
{\rm with}~~~ g_{\rm eff }&=\sum_{i=1}^{4} g_{i}\left(1+\Delta_{i}\right)^{3 / 2} e^{-x \Delta_{i}}\,,
\ea
\ee
where $x=m_1/T$, $\sigma_{i j}$ is the cross section for the reaction $\tilde{\Pi}^\pm_{i}\tilde{\Pi}^\mp_{j} \rightarrow \tilde{\Pi}^0\tilde{\Pi}^0$ given in \Cref{eq:thermalavcx}
(summed over all kinematically accessible SM pions in the final state), 
$g_{i}=2$ is the number of degrees of freedom of $\tilde{\Pi}^\pm_{i}$, and
$\Delta_{i} \equiv \left(m_{i}-m_{1}\right) / m_{1}$ is the mass difference between the heavier dark pions and $\tilde{\Pi}_1^\pm$. 

In \Cref{fig:sigmaeff} we present $\langle \sigma_{\rm eff} v\rangle$ for a range of $f$ and $m_{\rm DM}$. By fitting our numerical results to the approximation given in \Cref{eq:xsecapprox}, we find the velocity-averaged effective cross-section to be 
\begin{align}
\langle \sigma_{\rm eff} v\rangle \simeq (1.5-2)\times 10^{-11}\,{\rm GeV}^{-2}\left(\frac{m_{\rm DM}}{5~{\rm TeV}}\right)\left(\frac{65~{\rm TeV}}{f}\right)^3\,,
\end{align}
where the lower and higher values correspond to one or three generations of SM fermions respectively. For smaller $f$ and larger constituent DM mass, $\langle \sigma_{\rm eff} v\rangle$ is larger, resulting in too much annihilation and hence not enough dark pions left over to produce the observed abundance of the dark matter. 
Conversely, a lighter constituent dark matter mass and higher confinement scale result in a lower dark pion annihilation cross-section and an overabundance of dark matter.  

\subsection{Deconfinement}
The freeze-out of the dark pions determines the final comoving number density of dark pions, which has an associated energy density, $\rho_{\rm DP}=m_1n_{\rm DP}$. At the time of deconfinement, each dark pion flies apart into one $\chi$ as well as SM radiation. At that point, the
dark matter consists of freely streaming $\chi$ particles, with energy density,
\be
\label{eq:rhofinal}
\ba
\rho_{\rm DM} = \frac{m_{\rm DM}}{m_1} \times \rho_{\rm DP} = m_{\rm DM} \times n_{\rm DP}~,
\ea
\ee
which is to be compared with the observed abundance of dark matter from cosmological measurements, $\Omega h^2 = 0.1200 \pm 0.0012$
\cite{Aghanim:2018eyx}.

We assume that the weak sector deconfines at temperature $T_{\rm dc}$, where $T_{\rm dc}\sim m_{1}/100$.
In estimating the relic density of $\chi$, we assume that the entropy dump into the thermal plasma from the deconfinement process is negligible\footnote{The vacuum energy in the confined phase is $\sim c_0\Lambda_W^4$, where $c_0$ is a constant. 
We require that this energy is always smaller than the contribution from relativistic degrees of freedom in the Universe, $g_\ast T^4$. 
Assuming deconfinement happens at a temperature $T_{\rm dc}=10^{-4}\Lambda_W$, requiring $c_0\Lambda_W^4 < g_\ast T_{\rm dc}^4$ 
would imply that $c_0 \lesssim 10^{-14}$.}.
After deconfinement, the free $\chi$ particles could begin to annihilate into SM through the now unbound weak interactions, for which the
cross-section is parametrically $\sigma_{\rm W} \approx \alpha^2_{\rm W}\pi/m^2_{\rm DM}$, where $\alpha_{\rm W}\sim 0.1$ has presumably
returned to the value measured by experiments today. In our numerical scans, we verify that 
$\sigma_{\rm W} ~n_{\rm DP}\ll H$ at $x=100$ for the regions of $(m_{\rm DM},f)$ of interest, ensuring that no period of thermalization
after deconfinement occurs and therefore alters the dark matter relic density from \Cref{eq:rhofinal}.


\subsection{Numerical Results}

\begin{figure}[t!]
\centering
\includegraphics[width=0.49\textwidth]{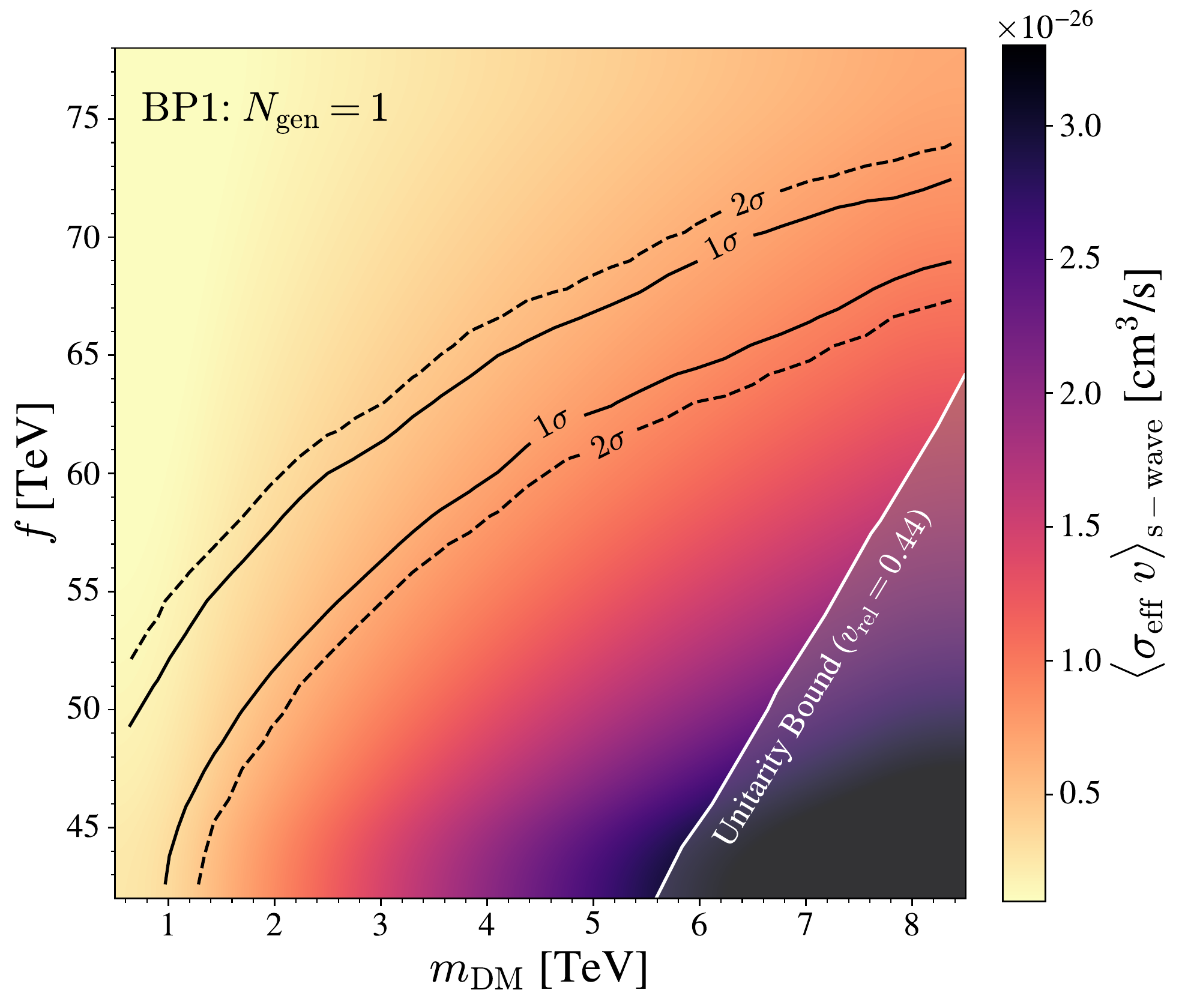}
\includegraphics[width=0.49\textwidth]{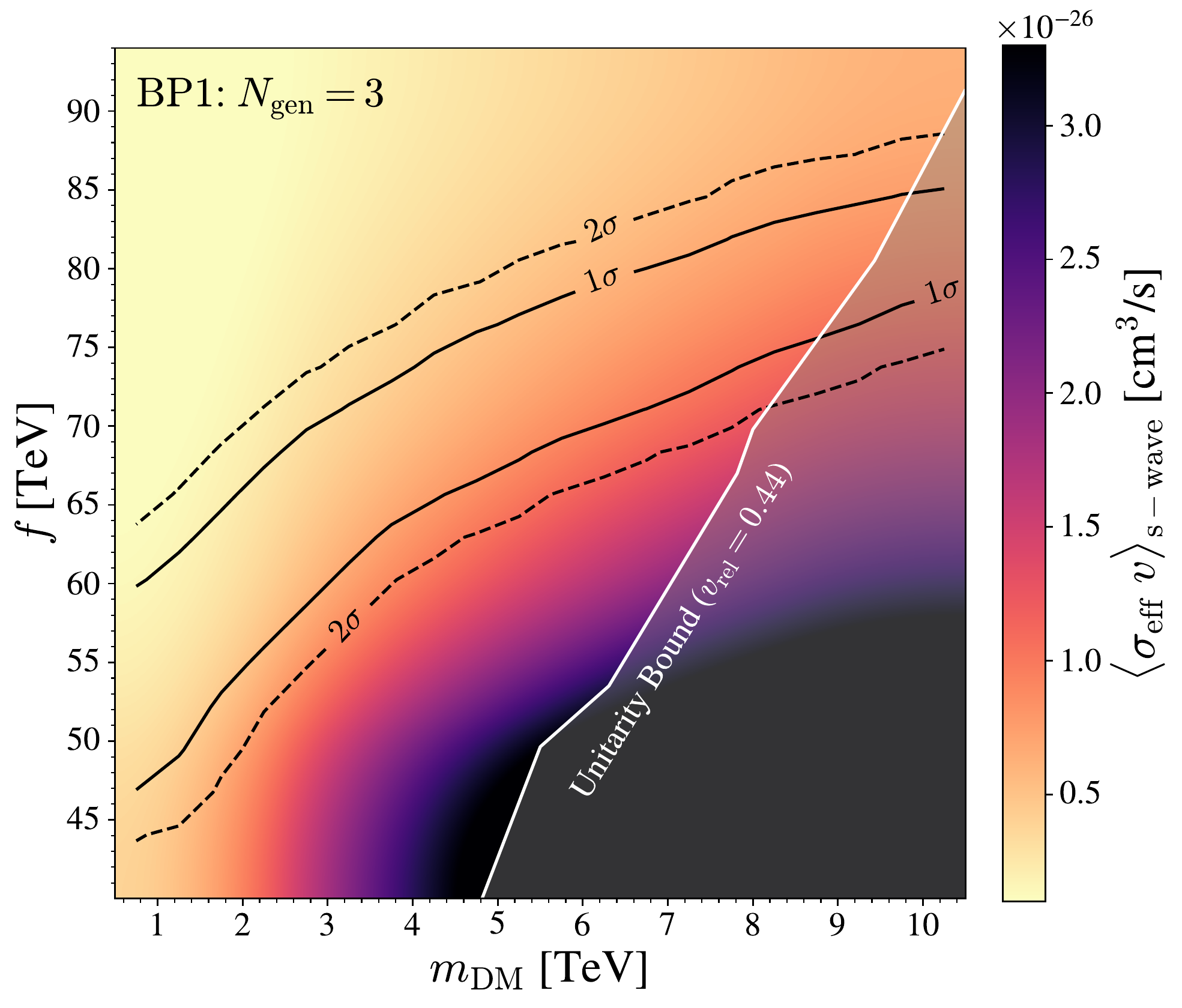}
\caption{The region of interest for the constituent dark matter mass, $m_{\rm DM}$, and the weak confinement scale, $f$, for one generation (left) and three generation (right) cases. The solid and dashed lines show where the DM relic density is consistent with observations at 1 and 2 $\sigma$ respectively. We show the velocity-averaged effective cross section during freeze-out given in \Cref{eq:sigmaeff}. The grey shaded area is inconsistent with unitarity constraints. Note that for both cases we start our scan at $m_{\rm DM}=500$~GeV and that the highest points for our scans are $m_{\rm DM}=8.5$~ TeV and 10.5 TeV for one generation and three generation case respectively. For the benchmark shown above, BP1, $g_{s} = 0.8, e_{Q} = 0.5$ and $s_Q^2 = 0.12$.}\label{fig:sigmaeff}
\end{figure}

We numerically solve \Cref{eq:DMbolzman}, adapting the infrastructure of {\sc ULYSSES} \cite{Granelli:2020pim}, a publicly available {\sc Python} 
package developed to solve Boltzmann equations associated with leptogenesis. 
For each benchmark point, we determine the regions of the parameter space, $(m_{\rm DM},f)$ that are consistent with the measured relic abundance. 
To perform this task, we use {\sc ULYSSES} in conjunction with {\sc
MultiNest}~\cite{Feroz:2008xx,Feroz:2007kg,2013arXiv1306.2144F} (more precisely,
{\sc pyMultiNest}~\cite{pymultinest}, a wrapper around {\sc Multinest} written
in {\sc Python}). We place flat priors on the parameters $(m_{\rm DM},f)$ and employ the log-likelihood as the {\sc Multinest} objective function:
\begin{equation}
\log L=-\frac{1}{2}\left(\frac{\Omega \text{h}^2(m_{\rm DM}, f)-\Omega \text{h}^2_{\rm{PDG}}}{\Delta \Omega \text{h}^2}\right)^{2}\,,
\end{equation}
where $\Omega \text{h}^2(m_{\rm DM},f)$ 
is the calculated relic density for a point in the model parameter space, $\Omega \rm{h}^2_{\rm{PDG}}$ is the best-fit value of the relic density and 
$\Delta \Omega \text{h}^{2}$ is the 1-$\sigma$ experimental uncertainty range of the relic abundance \cite{Aghanim:2018eyx}. 
In the left panel of \Cref{fig:sigmaeff}, 
we show the regions for which the predicted relic abundance of dark matter is consistent with the observed abundance at the one and two sigma. We find that
multi-TeV $\chi$ masses (and $f \sim 60$~TeV) are favored and consistent with the perturbative unitarity bound \cite{Griest:1989wd}, which,
using the approximate analytic form of $\langle \sigma_{\rm eff} v \rangle$ \Cref{eq:xsecapprox}, takes the form:
\begin{align}
\langle \sigma_{\rm eff} v \rangle_{\rm s-wave} \approx \frac{0.8 m_{\rm DM}}{f^3}\lesssim \frac{4\pi}{64\pi^3 f m_{\rm DM}v} \Rightarrow m_{\rm DM}^2\lesssim \frac{5f^2}{64\pi^2 v} \,,
\end{align}
where we substitute $m_{\rm DP}^2 = 64\pi^3fm_{\rm DM}$. For a freeze-out temperature of $T_{\rm fo}\simeq m_1/30$, the unitarity limit constrains $m_{\rm DM} \lesssim 1.3 f$, which cuts into the parameter regime favored by the relic density at around $m_{\rm DM}\sim 10$~TeV. 
\Cref{fig:sigmaeff} shows the unitarity limit on the region of interest using the numerical results for $\langle \sigma_{\rm eff} v \rangle_{\rm s-wave}$. 
The numerical results for BP1 and BP2 are qualitatively very similar. 
Our code, which calculates the effective cross-section and solves the Boltzmann equations, for both the one- and three-generation case, is publicly available at \href{https://github.com/jnhoward/SU2LDM_public}{\faGithub}.

\section{Three Generations of Standard Model doublets and Dark Matter}
\label{sec:3gen}

For simplicity, we have outlined the freeze-out dynamics in the case of a single generation of SM doublets together with the pair of vector-like fermionic $\SUw$  
doublets ($\{ \ell, q^r, q^g, q^b, \chi_1, \chi_2 \}$).
In this Section, we generalize to three generations ($\{ \ell_i, q_i^r, q_i^g, q_i^b, \chi_1, \chi_2 \}$ with $i = 1, 2, 3$) 
where there are 90 pseudo-Goldstone bosons and an $\eta^\prime$. The mass matrix is $91\times 91$ and, due to the added complexity of three generations of SM doublets, 
the mass$^2$ matrix contains off-diagonal entries which depend non-trivially on the scan parameters $(m_{\rm DM}, f)$. 
Therefore, unlike in the one generation case, where we could perform the 
diagonalization of the mass squared matrix analytically, in the three-generation case, we instead
rely on a numerical diagonalization of the mass-squared matrix to transform from the interaction to the mass basis for each parameter scan point. 
We perform the same procedure outlined in \secref{sec:wpm} to transform from the mass to the U(1)$_\chi$ basis, and compute annihilation cross-section as described in 
\secref{sec:annihilationCX}, but in the three-generation case, there are 12 charged dark pion states.
 We find that there are ten distinct pion masses as shown in \tabref{tab:pionMassesInt3}. 
 Rather than provide the complete indexing of states, we provide the number of pions (second column)
 with each mass eigenvalue. Interestingly, several new states, such as the color triplet, appear in the multi-generational case.
\begin{table}[t!]
\resizebox{\textwidth}{!}{%
\begin{tabular}{ |c|c|c|c|c|} 
\hline
Pion & $\#$ & Mass squared & $\U{\rm Q}$ & $SU(2)_C$ \\
(mass basis)& {} & value & charge & charge \\
\hline
\hline
 $\Pi_{1}^{\rm mass}$ &1 & $64 \pi ^2 f \left(7 f
 \kappa +\pi m_{\rm DM} + \sqrt{49 f^2 \kappa ^2-10 \pi f \kappa m_{\rm DM}+\pi ^2 m_{\rm DM}^2}\right)
$ & 0 & 1\\
\hline
$\Pi_{2}^{\rm mass}$ &24 & $-\frac{1}{2} C_A e_Q^2 f^2-\frac{3}{2} C_G f^2 g_s^2+C_W f^2
 g_s^2-\frac{1}{2} C_Z e_Q^2 f^2 s_Q^2+\frac{C_Z e_Q^2
 f^2}{6 s_Q^2}+\frac{1}{3} C_Z e_Q^2 f^2$ & $\pm 1$ & 2\\
\hline
$\Pi_{3}^{\rm mass}$ & 14 & $0$& $0$ & 1 \\
\hline
$\Pi_{4}^{\rm mass}$ & 6 &$-2 C_A e_Q^2 f^2-2 C_Z e_Q^2 f^2 s_Q^2-\frac{2 C_Z
 e_Q^2 f^2}{9 s_Q^2}+\frac{4}{3} C_Z e_Q^2 f^2$ & $\pm 1$ & 1\\
\hline
$\Pi_{5}^{\rm mass}$ & 12 & $-\frac{1}{2} C_A e_Q^2 f^2-\frac{3}{2} C_G f^2 g_s^2-C_W f^2
 g_s^2-\frac{1}{2} C_Z e_Q^2 f^2 s_Q^2+\frac{C_Z e_Q^2
 f^2}{6 s_Q^2}+\frac{1}{3} C_Z e_Q^2 f^2$ & $\pm 1$ & 2 \\
\hline
$\Pi_{6}^{\rm mass}$ & 6 & $64 \pi ^3 f m_{\rm DM}$ & $0$ & 1\\
\hline
$\Pi_{7}^{\rm mass}$ & 6 & $-2 C_A e_Q^2 f^2-2 C_Z e_Q^2 f^2 s_Q^2+\frac{2}{3} C_Z
 e_Q^2 f^2+64 \pi ^3 f m_{\rm DM}$ & $\pm 1$ & 1\\
 \hline
$\Pi_{8}^{\rm mass}$ & 9 &$-4 C_G f^2 g_s^2$ & 0 & 3\\
 \hline
 $\Pi_{9}^{\rm mass}$ &12 & $-\frac{1}{2} C_A e_Q^2 f^2-\frac{3}{2} C_G f^2 g_s^2-\frac{1}{2}
 C_Z e_Q^2 f^2 s_Q^2+\frac{C_Z e_Q^2 f^2}{18 s_Q^2}+64
 \pi ^3 f m_{\rm DM}$ & $\pm 1$ & 2\\
\hline
 $\Pi_{10}^{\rm mass}$ & 1& $64 \pi ^2 f \left(7 f
 \kappa +\pi m_{\rm DM}-\sqrt{49 f^2 \kappa ^2-10 \pi f \kappa m_{\rm DM}+\pi ^2 m_{\rm DM}^2}\right)$ & 0 & 1\\
\hline
\hline
\end{tabular}
}
\caption{Table of mass squared values corresponding to mass basis states along with the relevant $SU(2)_C\times \U{Q}$ charges. Three SM generations with $\chi_1$ and $\chi_2$ are included. }\label{tab:pionMassesInt3}
\end{table}

In the right panel of \Cref{fig:sigmaeff}, we show the regions for which the predicted relic abundance of dark matter 
in the three generation case is consistent with the observed abundance.
The favored region that explains the DM abundance in the three-generation case is approximately the same as the simplified one generation case but favors slightly higher $f$ values for a given $m_{\rm DM}$. Another slight difference is that the unitarity constraint is more stringent due to the higher values of $\langle \sigma_{\rm eff} v \rangle_{\rm s-wave}$ in the three-generation case.

\section{Outlook}
\label{sec:outlook}

Our results indicate that a modification to the strength of the $\SUw$ weak coupling dramatically transforms the nature of the freeze-out process for an $\SUw$-charged
WIMP. For a vector-like pair of doublets, we find that the
weak confinement scenario favors a range of masses (depending on the early $\SUw$ confinement scale) around $O(1-10)$ TeV and can be
much larger than the $\simeq 1.1$~TeV favored by a standard cosmological history \cite{Cirelli:2005uq}.  
This highlights the possibility that the physics of the dark matter itself could be drastically different at the time of freeze-out from today. In particular, 
the constraints on a several TeV WIMP are quite different from those restricting a $\sim$ 1 TeV mass particle.

Direct searches for WIMPs scattering with heavy nuclei remain an important challenge. At $\sim10$~TeV, XENON1T data restricts the cross-section to scatter with a nucleon
to be smaller than about $\sim 10^{-44}$~cm$^2$ \cite{XENON:2018voc}, which is still incompatible 
with the cross-section mediated by full strength $Z$ boson exchange ($\sim 10^{-38}$ cm$^2$). However, this bound can be avoided by introducing
Majorana masses via a dimension-5 operator of the form,
\be
\label{eq:majorana}
\ba
{\cal L}_{\Delta M} = \frac{1}{M_1} ( H^\dagger \chi_1) ( H^\dagger \chi_1) + \frac{1}{M_2} ( H \chi_2) ( H \chi_2) + \rm{h.c.}
\ea
\ee
where $M_{1,2}$ parameterize the interaction strength and
parentheses indicate how $\SUw$ indices are contracted.
After electroweak breaking, these operators result in Majorana masses of order $v^2 / M_{1,2}$, which split the Dirac $\chi$ into two Majorana fermions,
in close analogy with the see-saw mechanism for generating neutrino masses.
The Majorana particles have vanishing vector currents, and thus $Z$ boson exchange mediates inelastic scattering, which
is kinematically suppressed once the mass splitting is larger than the typical kinetic energy of the WIMPs in the Galactic halo \cite{Tucker-Smith:2001myb,Bramante:2016rdh}.
Provided the scales $M_1$ and $M_2$ are sufficiently large, these operators play essentially no role in freeze-out, and do not themselves mediate an observable
scattering with nuclei via Higgs boson exchange.

Despite its full-strength electroweak interactions, a multi-TeV dark matter particle is too heavy to be accessible at the LHC. 
Even when kinematically accessible, unless there 
is mixing with another nearby state via electroweak symmetry-breaking, the signatures at colliders are challenging because the charged state is expected to be
degenerate with its neutral counter-part to within a few hundred MeV \cite{Hill:2011be}, and thus requires mono-jet or disappearing track analyses.
As a result, even a future 100 TeV hadron collider is expected to struggle to reach sensitivity to TeV mass electroweak doublets \cite{Low:2014cba}.

Indirect searches for the annihilation products of WIMP annihilation, for example, from observation of high energy $\gamma$-rays, can reach sensitivity
to around 10 TeV for electroweak-sized annihilation cross-sections \cite{MAGIC:2016xys}, 
particularly for masses for which the annihilation experiences a Sommerfeld-like enhancement due to the exchange of weak bosons. These bounds exhibit
a considerable sensitivity to the distribution profile of the dark matter 
around the Galactic center, which is not well constrained by observation (see, e.g. Ref.~\cite{Arakawa:2021vih} for discussion).
Despite these challenges, a future gamma-ray observatory such as the Cherenkov Telescope Array \cite{CTA:2020qlo} could offer the best chance of a direct
observation of dark matter in such a scenario.

Looking forward, it would be interesting to explore further the consequences of a period of early $\SUw$ confinement. It may be that such an epoch could enable new possibilities
to understand other mysteries of the early Universe, such as the primordial asymmetry between baryons and anti-baryons. And more widely, our results illustrate
the general truth that the early Universe may well turn out to have been more weird and wonderful than simply extrapolating the SM to high temperatures
would lead us to expect. Exploring the space of possibilities and how to constrain them with experimental measurements will remain an essential task for particle physics.

\section*{Acknowledgements}
We are grateful to Holger Schulz for his insightful advice regarding speeding up our Boltzmann code and
to Daniel Whiteson for his insight into LHC collider constraints.
We  thank Joshua Berger and Andrew Long for their helpful comments on the draft. 
This work was supported in part by the National Science Foundation via grant numbers PHY-1915005 and DGE-1839285. Any opinions, findings and conclusions or recommendations expressed in this material are those of the author(s) and do not necessarily reflect the views of the National Science Foundation.
This work used the DiRAC@Durham facility managed by the Institute for Computational Cosmology on behalf of the STFC DiRAC HPC Facility (www.dirac.ac.uk). The equipment was funded by BEIS capital funding via STFC capital grants ST/P002293/1, ST/R002371/1 and ST/S002502/1, Durham University and STFC operations grant ST/R000832/1. DiRAC is part of the National e-Infrastructure.

\appendix

\section{Matrices for One Generation}
\label{app:A}

The explicit form of the matrices in \equaref{eq:potential1} for a single generation of Standard Model doublets in addition to $\chi_1$ and $\chi_2$ are:
\be
\ba
L_1 &= 
\left(
\begin{array}{cccccc}
 0 & 0 & 0 & 0 & 0 & 0 \\
 0 & 0 & 0 & 0 & 0 & 0 \\
 0 & 0 & 0 & \frac{1}{2} & 0 & 0 \\
 0 & 0 & \frac{1}{2} & 0 & 0 & 0 \\
 0 & 0 & 0 & 0 & 0 & 0 \\
 0 & 0 & 0 & 0 & 0 & 0 \\
\end{array}
\right), & L_2 &= \left(
\begin{array}{cccccc}
 0 & 0 & 0 & 0 & 0 & 0 \\
 0 & 0 & 0 & 0 & 0 & 0 \\
 0 & 0 & 0 & -\frac{i}{2} & 0 & 0 \\
 0 & 0 & \frac{i}{2} & 0 & 0 & 0 \\
 0 & 0 & 0 & 0 & 0 & 0 \\
 0 & 0 & 0 & 0 & 0 & 0 \\
\end{array}
\right),
 & L_{3} &= \left(
\begin{array}{cccccc}
 0 & 0 & 0 & 0 & 0 & 0 \\
 0 & 0 & 0 & 0 & 0 & 0 \\
 0 & 0 & \frac{1}{2} & 0 & 0 & 0 \\
 0 & 0 & 0 & -\frac{1}{2} & 0 & 0 \\
 0 & 0 & 0 & 0 & 0 & 0 \\
 0 & 0 & 0 & 0 & 0 & 0 \\
\end{array}
\right),
\\
Q &= \left(
\begin{array}{cccccc}
 \frac{1}{2} & 0 & 0 & 0 & 0 & 0 \\
 0 & -\frac{1}{2} & 0 & 0 & 0 & 0 \\
 0 & 0 & 0 & 0 & 0 & 0 \\
 0 & 0 & 0 & 0 & 0 & 0 \\
 0 & 0 & 0 & 0 & \frac{1}{2} & 0 \\
 0 & 0 & 0 & 0 & 0 & -\frac{1}{2} \\
\end{array}
\right), & L^{1, +} &= \left(
\begin{array}{cccccc}
 0 & 0 & 0 & 0 & 0 & 0 \\
 0 & 0 & 1 & 0 & 0 & 0 \\
 0 & 0 & 0 & 0 & 0 & 0 \\
 0 & 0 & 0 & 0 & 0 & 0 \\
 0 & 0 & 0 & 0 & 0 & 0 \\
 0 & 0 & 0 & 0 & 0 & 0 \\
\end{array}
\right), & L^{2, +}& =\left(
\begin{array}{cccccc}
 0 & 0 & 0 & 0 & 0 & 0 \\
 0 & 0 & 0 & 1 & 0 & 0 \\
 0 & 0 & 0 & 0 & 0 & 0 \\
 0 & 0 & 0 & 0 & 0 & 0 \\
 0 & 0 & 0 & 0 & 0 & 0 \\
 0 & 0 & 0 & 0 & 0 & 0 \\
\end{array}
\right), \\
L^{1, -} &= \left(
\begin{array}{cccccc}
 0 & 0 & 0 & 0 & 0 & 0 \\
 0 & 0 & 0 & 0 & 0 & 0 \\
 0 & 1 & 0 & 0 & 0 & 0 \\
 0 & 0 & 0 & 0 & 0 & 0 \\
 0 & 0 & 0 & 0 & 0 & 0 \\
 0 & 0 & 0 & 0 & 0 & 0 \\
\end{array}
\right) , & L^{2, -} &=\left(
\begin{array}{cccccc}
 0 & 0 & 0 & 0 & 0 & 0 \\
 0 & 0 & 0 & 0 & 0 & 0 \\
 0 & 0 & 0 & 0 & 0 & 0 \\
 0 & 1 & 0 & 0 & 0 & 0 \\
 0 & 0 & 0 & 0 & 0 & 0 \\
 0 & 0 & 0 & 0 & 0 & 0 \\
\end{array}
\right) ,
 & M&=\frac{m_{\rm DM}}{2} \begin{pmatrix}
 0 & 0 & 0 & 0 & 0 & 0 \\
 0 & 0 & 0 & 0 & 0 & 0 \\
 0 & 0 & 0 & 0 & 0 & 0 \\
 0 & 0 & 0 & 0 & 0 & 0 \\
 0 & 0 & 0 & 0 & 0 &1\\
 0 & 0 & 0 & 0 & -1 & 0 \\
\end{pmatrix} , \\
 J&=
\left(
\begin{array}{cccccc}
 -\frac{s^2_{Q}}{2} & 0 & 0 & 0 & 0 & 0 \\
 0 & \frac{s^2_{Q}}{2}-\frac{1}{3} & 0 & 0 & 0 & 0 \\
 0 & 0 & \frac{1}{6} & 0 & 0 & 0 \\
 0 & 0 & 0 & \frac{1}{6} & 0 & 0 \\
 0 & 0 & 0 & 0 & -\frac{s^2_{Q}}{2} & 0 \\
 0 & 0 & 0 & 0 & 0 & \frac{s^2_{Q}}{2} \\
\end{array}
\right) .
\ea
\ee

\bibliographystyle{JHEP}
\bibliography{references}
\end{document}